\documentclass{nature}

 \makeatletter
\def\@fnsymbol#1{\ensuremath{\ifcase#1\or \dagger\or \ddagger\or
   \mathsection\or \mathparagraph\or \|\or **\or \dagger\dagger
   \or \ddagger\ddagger \else\@ctrerr\fi}}
    \makeatother

\usepackage{xcolor}
\usepackage{comment}
\usepackage[utf8]{inputenc}
\usepackage{amsmath, amsthm, amssymb} 

\usepackage{lineno}

\usepackage{ulem}

\makeatletter
\renewcommand*{\@fnsymbol}[1]{\ensuremath{\ifcase#1\or \dagger\or *\or \ddagger\or
   \mathsection\or \mathparagraph\or \|\or **\or \dagger\dagger
   \or \ddagger\ddagger \else\@ctrerr\fi}}
\makeatother

\title{First experimental search for production of magnetic monopoles via the Schwinger mechanism}

\author{B. Acharya$^{1}$, J. Alexandre$^{1}$, P. Benes$^{2}$, B. Bergmann$^{2}$, S. Bertolucci$^{3}$, A. Bevan$^{4}$, H. Branzas$^{5}$, P.~Burian$^{2}$, M. Campbell$^{6}$, Y. M. Cho$^{7}$, M. de Montigny$^{8}$, A. De Roeck$^{6}$, J.~R.~Ellis$^{1,9}$, M. El Sawy$^{6}$, M. Fairbairn$^{1}$, D. Felea$^{5}$, M. Frank$^{10}$, O. Gould$^{11,12}$, J. Hays$^{4}$, A. M. Hirt$^{13}$, D.~L.~J.~Ho$^{14}$, P.~Q.~Hung$^{15}$, J. Janecek$^{2}$, M. Kalliokoski$^{16}$, A. Korzenev$^{17}$, D. H. Lacarr\`{e}re$^{6}$, C. Leroy$^{18}$, G.~Levi$^{19}$, A.~Lionti$^{17}$, A. Maulik$^{3,8}$, A. Margiotta$^{19}$, N. Mauri$^{3}$, N. E. Mavromatos$^{1,20}$, P. Mermod$^{17,\thanks{deceased}}$, L. Millward$^{4}$, V. A. Mitsou$^{21}$, I. Ostrovskiy$^{22,\thanks{corresponding author, iostrovskiy@ua.edu}}$, P.-P. Ouimet$^{8}$, J.~Papavassiliou$^{21}$, B. Parker$^{23}$, L. Patrizii$^{3}$, G. E. P\u{a}v\u{a}la\c{s}$^{5}$, J. L. Pinfold$^{8}$, L. A. Popa$^{5}$, V. Popa$^{5}$, M.~Pozzato$^{3}$, S. Pospisil$^{2}$, A. Rajantie$^{14}$, R. Ruiz de Austri$^{21}$, Z. Sahnoun$^{3}$, M. Sakellariadou$^{1}$,
A.~Santra$^{21}$, S. Sarkar$^{1}$, G. Semenoff$^{24}$, A. Shaa$^{8}$, G. Sirri$^{3}$, K. Sliwa$^{25}$, R. Soluk$^{8}$, M. Spurio$^{19}$, M.~Staelens$^{8}$,
M. Suk$^{2}$, M. Tenti$^{26}$, V. Togo$^{3}$, J. A. Tuszy\'{n}ski$^{8}$, A. Upreti$^{22}$, V. Vento$^{21}$, O. Vives$^{21}$}

\begin{document}

\maketitle

\begin{affiliations}
\item Theoretical Particle Physics \& Cosmology Group, Physics Dept., King's College London, UK
\item IEAP, Czech Technical University in Prague, Czech Republic
\item INFN, Section of Bologna, Bologna, Italy
\item School of Physics and Astronomy, Queen Mary University of London, UK
\item Institute of Space Science, Bucharest - M\u{a}gurele, Romania
\item Experimental Physics Department, CERN, Geneva, Switzerland
\item Center for Quantum Spacetime, Sogang University, Seoul, Korea
\item Physics Department, University of Alberta, Edmonton, Alberta, Canada
\item Theoretical Physics Department, CERN, Geneva, Switzerland
\item Department of Physics, Concordia University, Montr\'{e}al, Qu\'{e}bec, Canada
\item Physics Department, University of Helsinki, Helsinki, Finland
\item University of Nottingham, Nottingham, UK
\item Department of Earth Sciences, Swiss Federal Institute of Technology, Zurich, Switzerland
\item Department of Physics, Imperial College London, UK
\item Department of Physics, University of Virginia, Charlottesville, VA, USA
\item Helsinki Institute of Physics, University of Helsinki, Helsinki, Finland
\item D\'{e}partement de Physique Nucl\'{e}aire et Corpusculaire, Universit\'{e} de Gen\`{e}ve, Geneva, \\ Switzerland
\item D\'{e}partement de Physique, Universit\'{e} de Montr\'{e}al, Qu\'{e}bec, Canada
\item INFN, Section of Bologna \& Department of Physics \& Astronomy, University of Bologna, Italy
\item Department of Physics, School of Applied Mathematical and Physical Sciences, National Technical University of Athens, Zografou Campus, Athens GR-157 80 Greece
\item IFIC, Universitat de Val\`{e}ncia - CSIC, Valencia, Spain
\item Department of Physics and Astronomy, University of Alabama, Tuscaloosa, Alabama, USA
\item Institute for Research in Schools, Canterbury, UK 
\item Department of Physics, University of British Columbia, Vancouver, British Columbia, Canada
\item Department of Physics and Astronomy, Tufts University, Medford, Massachusetts, USA
\item INFN, CNAF, Bologna, Italy
\end{affiliations}

\begin{abstract}
Schwinger showed~\cite{schwinger} that electrically-charged particles can be produced in a strong electric field by quantum tunnelling through the 
Coulomb barrier. By electromagnetic duality, if magnetic monopoles (MMs) exist, they would be produced by the same mechanism in a sufficiently strong magnetic field~\cite{affleck}. Unique advantages of the Schwinger mechanism are that its rate can be calculated using semiclassical techniques without relying on perturbation theory, and the finite MM size and strong MM-photon coupling are expected to enhance their production. Pb-Pb heavy-ion collisions at the LHC produce the strongest known magnetic fields in the current Universe~\cite{Huang_2016}, and this article presents the first search for MM production by the Schwinger mechanism. It was conducted by the MoEDAL experiment during the 5.02~{\rm TeV}/nucleon heavy-ion run at the LHC in November 2018, during which the MoEDAL trapping detectors (MMTs) were exposed to 0.235 nb$^{-1}$ of Pb-Pb collisions. The MMTs were scanned for the presence of magnetic charge using a SQUID magnetometer. MMs with Dirac charges 1$g_D$ $\leq$ $g$ $\leq$ 3$g_D$ and masses up to 75 GeV/c$^2$ were excluded by the analysis. This provides the first lower mass limit for finite-size MMs from a collider search and significantly extends previous mass bounds. 
\end{abstract}

Magnetic monopoles (MMs) are hypothetical 
fundamental particles that carry isolated magnetic charge. Dirac formulated the first field theory of a point-like magnetic charge interacting with quantum charged matter and showed~\cite{Dirac:1931} that the existence of a MM would necessitate the quantization of electric charge. MMs also appear as solutions of Grand Unified Theories (GUT)~\cite{Hooft:1974,Polyakov:1974ek}. In contrast to the Dirac monopole, GUT MMs are composites of the fundamental non-Abelian gauge and Higgs fields that characterise the theory. Their expected masses are close to the GUT scale ($10^{16}$ GeV/c$^2$) and thus cannot be produced in a realistic collider experiment. Moreover, GUT monopoles are expected to have been diluted by cosmic inflation, and hence it is unlikely that they will be observed even in cosmic ray searches. String theories also predict MMs, which may have masses significantly lower than the GUT scale, depending on the string scale. In recent years composite finite-energy MM solutions of field theories beyond the Standard Model were also discovered in various contexts, possibly with masses as low as a few TeV~\cite{Mavromatos:2020gwk}. This raises the prospects for collider production of such objects, thus reviving interest in experimental searches for MMs.

All searches for the direct production of MMs at particle accelerators to date have focused on collisions of elementary particles such as electrons, or quarks in the case of hadron collisions, assuming production via fermion-antifermion annihilation (the Drell-Yan mechanism) or photon-photon collisions. However, the strong coupling of the MM~\cite{Dirac:1931} makes it difficult to calculate the production cross section. Indeed, it has been argued that the production of composite monopoles from elementary particle collisions is exponentially suppressed by $e^{-4/\alpha}$, where $\alpha$ is the electromagnetic fine structure constant at the energy scale set by the MM mass~\cite{Witten:1979kh,drukier}. However, the arguments for exponential suppression are based on perturbative considerations, and it is not known what a purely nonperturbative treatment, which is still lacking, could imply. Collider experiments have nevertheless focused on interpreting searches in the context of point-like MMs~\cite{Blagojevic:1985sh}, whereas most MMs predicted by models considered in the literature are composite objects, including the light MMs potentially accessible at the LHC~\cite{Mavromatos:2020gwk}. 

A way to avoid these shortcomings is provided by searching for MM pair production via the Schwinger mechanism. In 1951, Schwinger~\cite{schwinger} showed that electrically-charged particles could be produced in a strong electric field by tunnelling through the Coulomb barrier. If MMs exist, electromagnetic duality, supported by explicit calculations~\cite{affleck,Ho:2021uem}, implies that they would also be produced by the same mechanism in a sufficiently strong magnetic field. Unprecedented magnetic fields are present, fleetingly, in ultraperipheral heavy-ion collisions~\cite{Huang_2016}. In this case, the MM production cross section can be computed without relying on the perturbation theory (i.e., nonperturbatively) using semiclassical techniques, including the effects of strong MM-photon coupling. Due to the coherence of the magnetic field, the potential exponential suppression by $e^{-4/\alpha}$ is absent for MM production through the Schwinger mechanism. Additionally, in this approach, the finite size and strong coupling of MMs enhances their production~\cite{Ho:2019ads,Ho:2021uem}.

Although Schwinger production of MMs in a constant, or slowly-varying, magnetic field is well understood theoretically, and the production probability has been calculated accurately from first principles~\cite{affleck,Ho:2021uem}, the strong space and time dependence of the electromagnetic fields of LHC heavy-ion collisions present additional theoretical challenges. Progress on this front is made possible due to the large charges of heavy ions, as a consequence of which the resulting electromagnetic field can be approximated as a coherent classical field sourced by the ions' charge distribution. The strongest fields are generated in ultraperipheral collisions, for which the impact parameter is approximately twice the nuclear radius. In the 2018 heavy-ion run at the LHC, the peak magnetic field strength was~\cite{arttu_prd2019} \(B\approx 10^{16}~{\rm T}\), with an inverse decay time \(\omega\approx 10^{26}~{\rm s^{-1}}\). This field strength is about seven orders of magnitude greater than the critical field strength of quantum electrodynamics, and more than four orders of magnitude higher than the strongest known astrophysical magnetic fields, which are present on the surfaces of magnetars~\cite{Kaspi:2017fwg}.

We have considered two approximate approaches to the calculation of the overall MM production cross section: the free-particle approximation (FPA) given by equation~(\ref{equ:FPA}) in the Methods section and the locally-constant field approximation (LCFA) given by equation~(\ref{equ:LCFA}). In the FPA, the spacetime dependence of the electromagnetic field of the heavy ions is treated exactly, but MM self-interactions are neglected. Conversely, in the LCFA, the spacetime dependence of the electromagnetic field is neglected but MM self-interactions are treated exactly. In this way the two approximations are complementary, with uncorrelated uncertainties. In addition, for the FPA, the leading effects of MM self-interactions have been shown to enhance the cross section, and for the LCFA the leading effects of spacetime dependence have also been shown to enhance the cross section~\cite{arttu_prd2019,Gould:2021bre}. Thus, while neither approximation provides a complete calculation of the production cross section, both are expected to yield conservative lower limits. We conservatively use the smaller cross section in our final MM mass bounds.

In order to study experimentally the production of MMs created via the Schwinger mechanism in Pb-Pb heavy-ion collisions at the LHC, one also needs to be able to calculate the kinematics of the produced MMs, thus allowing the efficiency of their detection to be assessed. At LHC energies, the expected momentum distribution of MMs is primarily due to the time dependence of the electromagnetic field of the ultrarelativistic heavy ions. The momentum distribution predicted by the LCFA is narrower than what is allowed by the time-energy uncertainty principle, whereas the FPA prediction saturates it~\cite{Gould:2021bre}. Thus, the LCFA is a less reliable approximation for calculating the kinematics of Schwinger-produced MMs compared to the FPA, so only the latter is used in this work. The expression for this is given in equation~(\ref{equ:relFPA}) in the Methods section.

\begin{figure}[htpb]
\centering
\includegraphics[width=0.9\textwidth]{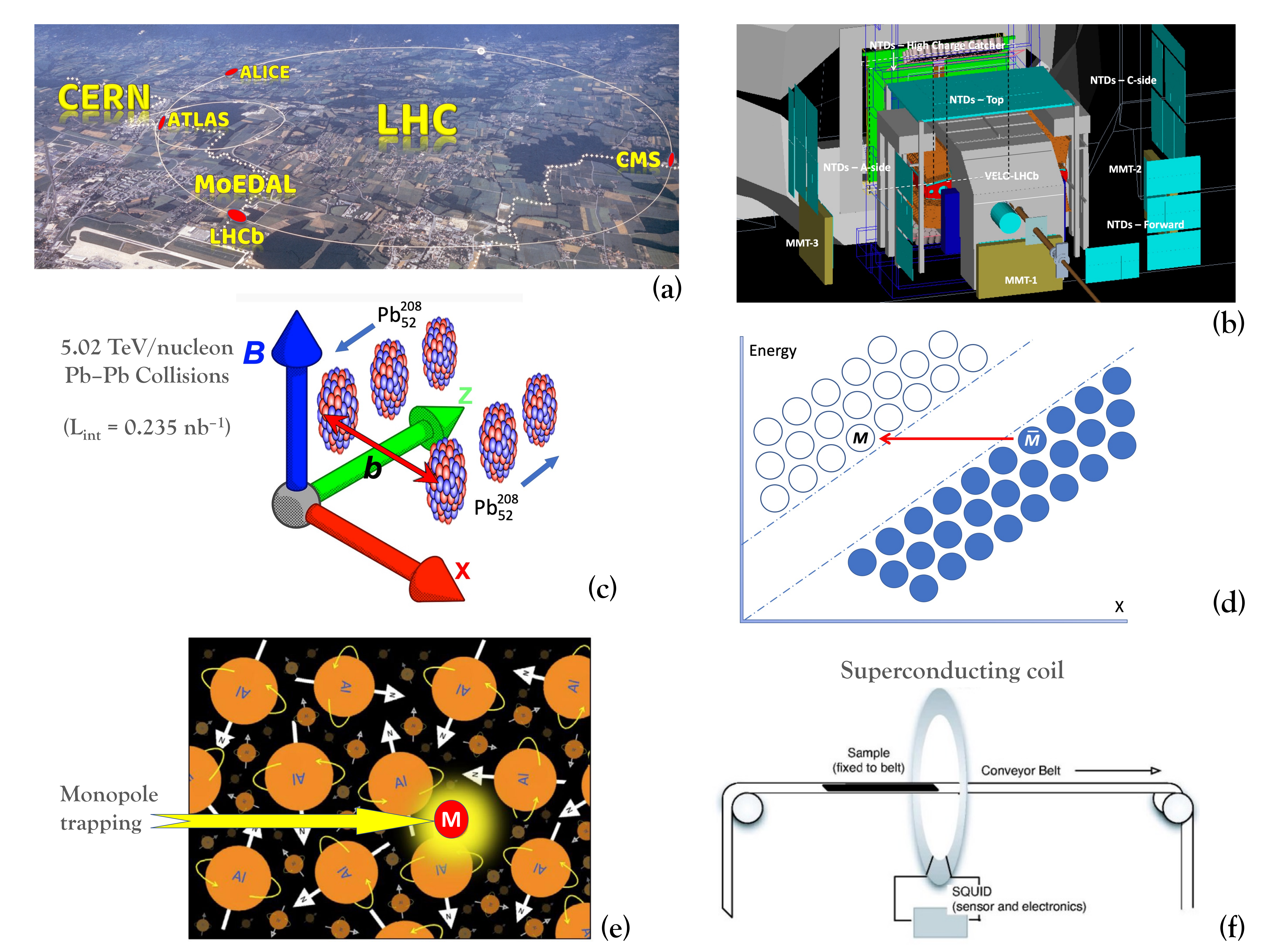}
\caption{Schematic diagram of the search. (a) The MoEDAL experiment is located at the Interaction Point 8 of the LHC. (b) It has an array of MMT detectors around the Interaction Point. (c) Peripheral Pb-Pb heavy-ion collisions produce strong magnetic fields. (d) These may produce magnetic monopole-antimonopole pairs via tunneling through the potential barrier (the Schwinger mechanism). (e) After production a magnetic monopole may be trapped in an MMT detector. (f) Samples from the MMTs are passed through a superconducting coil, and the magnetic charge of a trapped magnetic monopole will induce a signal in a SQUID detector.} 
\label{fig:diagram}
\end{figure}

The Monopole and Exotics Detector at the LHC (MoEDAL) is optimized for the detection of MMs and other highly-ionizing particles. MoEDAL has reported one of the strongest limits on point-like MM production in proton-proton collisions~\cite{moedal_prl2019}, and has set more recently first limits on the production of dyons (particles with both magnetic and electric charges) in a collider experiment~\cite{Dyons:2020}. Deployed in the LHCb VELO cavern around intersection point IP8, MoEDAL comprises two independent, passive detectors -- the Magnetic Monopole Trappers (MMTs) and Nuclear Track detectors (NTDs). The MMTs consist of 800 kg of aluminium blocks deployed in the forward and lateral regions around IP8. Because of its large anomalous magnetic moment, it is estimated that a $^{27}_{13}$Al nucleus (100\% natural abundance) would bind a magnetically-charged particle with an energy of 0.5–2.5 MeV~\cite{Milton:2006,Moedal:2014,Gamberg:2000}. After exposure, the MMT blocks are scanned for trapped magnetic charges with a SQUID magnetometer at the ETH Zurich Laboratory for Natural Magnetism. Figure~\ref{fig:diagram} shows a schematic diagram of this search. 

The mean expected rate of trapped MMs, $R_{exp}$, is defined as the mean number of MMs expected to be trapped during the Pb-Pb collision run. It is determined using a Monte Carlo (MC) simulation that is described further in the Methods section. For each MM mass and magnetic charge considered, the initial momenta are sampled from the Schwinger FPA kinematic distribution. The MMs are then propagated through the realistic geometry of the LHCb and MoEDAL detectors using the \textsc{Geant4} toolkit~\cite{geant4}. 
The MMT trapping efficiency, $\epsilon$, is defined as the ratio of the number of MMs trapped by MMTs to the total number of generated MMs. The mean expected rate is calculated for a given mass and magnetic charge as a product of the trapping efficiency, the luminosity, and the production cross section. 

The systematic uncertainty in the expected rate of trapped MMs due to the finite statistics of the MC samples is $<$ 2\% for magnetic charge of 1 $g_D$, increasing to 2-7\% for 2$g_D$ and reaching 20\% for 3$g_D$. The systematics in the calculation of energy losses by \textsc{Geant4} are between 1-7 \%~\cite{moedal_first}, and the uncertainty in the luminosity is better than 5\%. The dominant systematic uncertainty arises from the efficiency calculation, and is mainly due to the uncertainty in the material budget, which is determined by inspection and direct measurement. To estimate the related uncertainty, two material geometry models -- the minimal and maximal geometry variants -- are utilized as conservative limiting cases. For magnetic charges larger than 3$g_D$ all simulated MMs range out in the default geometry before reaching the MMTs. A more detailed discussion of the uncertainty calculation can be found in the Methods section.

\section*{Results}
All the MMT volumes exposed to 0.235 nb$^{-1}$ of 5.02 TeV per nucleon Pb-Pb collisions obtained in November 2018~\footnote{The much smaller (less than 10 $\mu$b$^{-1}$) Pb-Pb run in 2016 was not included in the analysis, in order to ensure uniform beam conditions.} were scanned using the DC-SQUID magnetometer for the presence of trapped magnetic charges. No statistically significant signal was observed. The existence of a MM with $\mid g\mid$ $\geq$ 0.5$g_{D}$ in the trapping volume was excluded at more than 3$\sigma$.

Comparing with the calculated production cross section, we exclude MMs with masses up to 75 GeV/c$^2$ to 70 GeV/c$^2$ for magnetic charges from 1$g_D$ to 3$g_D$, respectively, as shown in detail in Table~\ref{tab:masslimits}. The statistical significance of the limits take into account the uncertainty of $R_{exp}$ due to the systematics (see Tables~\ref{tab:nexpFPA1}\textendash~\ref{tab:nexpFPA5} in the Methods section) and Poisson statistics. 
\begin{table}[htpb]
\centering
\caption{95\% C.L. mass limits (GeV/c$^2$) on MM pair production in LHC Pb-Pb collisions}
\vspace{0.5cm}
\label{tab:masslimits}
\setlength{\tabcolsep}{12pt} 
\renewcommand{\arraystretch}{1} 
\begin{tabular}{l|c|c|c}
\hline
\multirow{2}{*}{} Cross section approximation & \multicolumn{3}{c}{Magnetic Charge ($g_{D}$)} \\ \cline{2-4}
& 1 & 2 & 3 \\ \hline
FPA & 90 & 70 & 70 \\ \hline
LCFA & 75 & 210 & 388 \\ \hline
Conservative limit & \textbf{75} & \textbf{70} & \textbf{70} \\ \hline
\end{tabular}
\end{table}
The expected numbers of events decrease rapidly with increasing MM mass, so the limits are relatively insensitive to the systematic uncertainties. Details of the measurement, calibration and statistical procedures are provided in the Methods section. We note that the FPA mass bounds for $|g| \ge 2 g_D$ are lower than for 1 $g_D$, because the increase in the cross section is offset by the rapid decrease in efficiency for higher magnetic charges. This is not the case for the LCFA, where the cross section grows for higher magnetic charges faster than the efficiency drops.

\begin{figure}[htpb]
\centering
\includegraphics[width=0.6\textwidth]{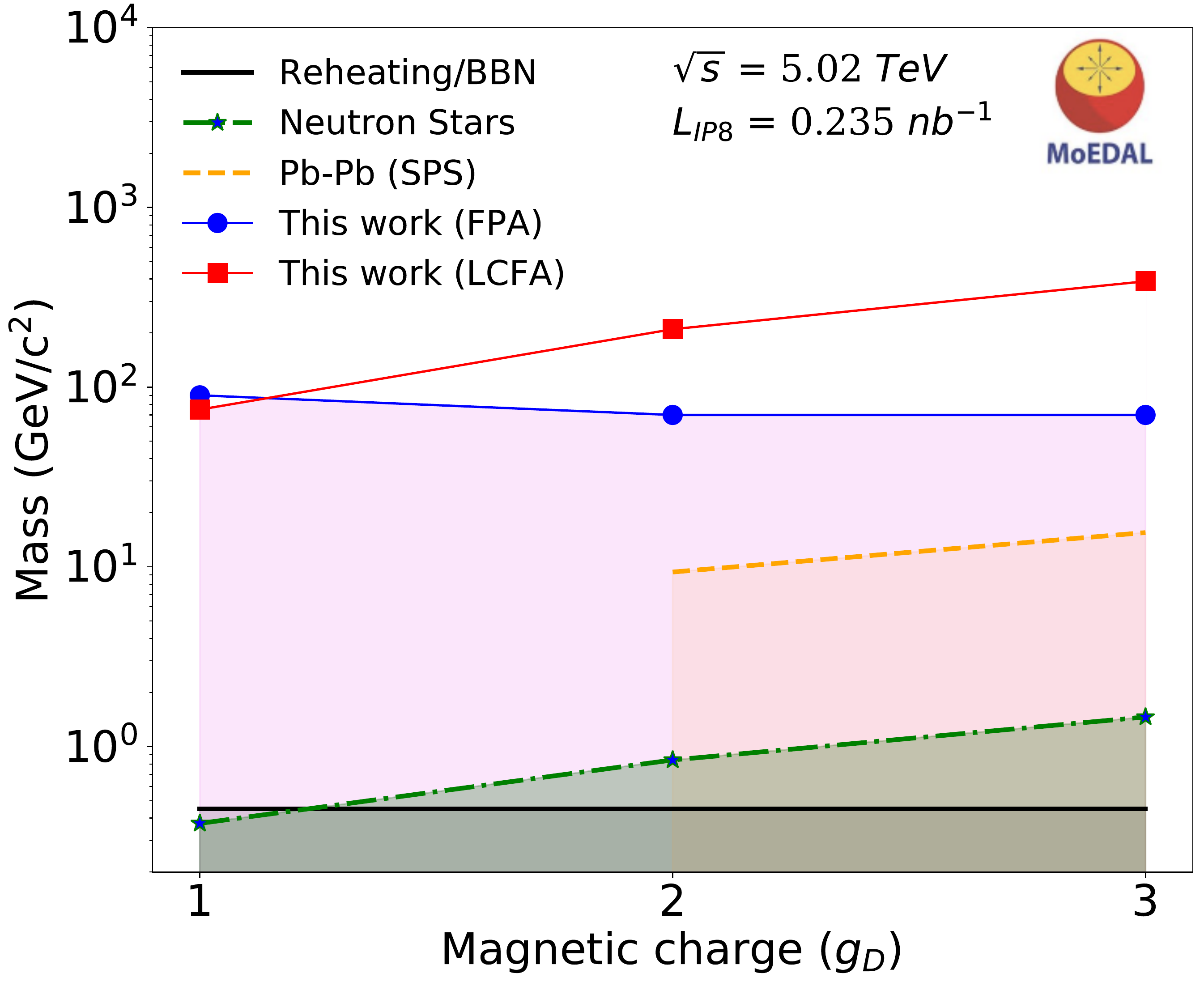}
\caption{The 95\% C.L. exclusion regions obtained using the FPA (blue) and LCFA (red) calculations of MM production via the Schwinger mechanism for 0.235 nb$^{-1}$ of 5.02 TeV per nucleon Pb-Pb collisions, with the conservative exclusion region shaded violet. Limits resulting from alternative production channels~\cite{arttu_prl2017} are also shown for comparison.}
\label{fig:MassBounds}
\end{figure}

Figure~\ref{fig:MassBounds} shows the exclusion regions in the magnetic charge vs. mass plane for the FPA and LCFA approximations. Our results exclude a significantly larger region of the parameter space than previous limits from alternative production channels, including production in heavy-ion collisions at the CERN SPS, interpreted in Ref.~\cite{arttu_prl2017} in terms of the Schwinger mechanism. Figure~\ref{fig:MassBounds} reports also limits from indirect searches of MMs produced close to the surface of neutron stars, as well as from cosmological reheating and Big-Bang Nucleosynthesis~\cite{arttu_prl2017}.

\begin{figure}[htpb]
\centering
\includegraphics[width=0.6\textwidth]{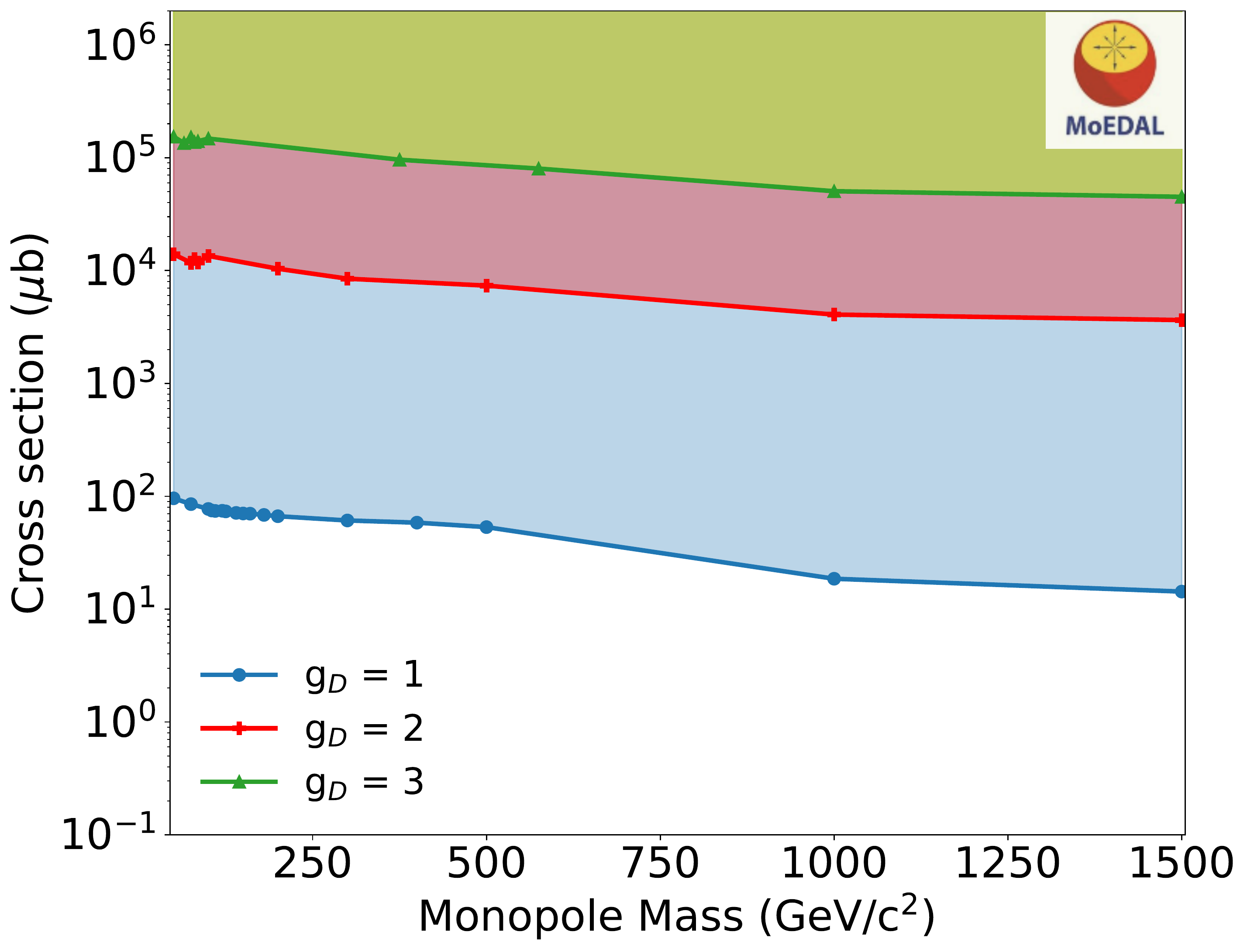}
\caption{The 95\% C.L. exclusion regions on the cross section for MM production via the Schwinger mechanism for 0.235 nb$^{-1}$ of 5.02 TeV per nucleon Pb-Pb collisions, as functions of the MM mass for magnetic charges 1$g_D$ (blue), 2$g_D$ (red), and 3$g_D$ (green).} 
\label{fig:exptXsection}
\end{figure}

The main result of this work (Figure~\ref{fig:MassBounds}) takes advantage of the recent progress in the calculation of the overall production rate of MMs in strong magnetic fields, which allows one to place limits on MM masses. An alternative way to present the results is to interpret the zero observed rate of MMs as a first experimental constraint on the cross section for the Schwinger mechanism for MM production, assuming the FPA production kinematics. This approach is complementary and free of potential theoretical prejudice on the overall production rate, relying only on the observed rate, measured luminosity, and MM trapping efficiency calculated under the FPA assumption. Figure~\ref{fig:exptXsection} shows the 95\% exclusion regions on the cross section for magnetic charges 1, 2, and 3 $g_D$.

\section*{Conclusions}
The MoEDAL detector took 0.235 nb$^{-1}$ of data in the 5.02 TeV per nucleon Pb-Pb heavy-ion run that took place in November 2018, the last year of LHC's Run-2. The MMT detectors exposed during this run were scanned for the presence of trapped magnetic charge using the SQUID magnetometer. No signal candidates were observed. This allowed the exclusion of production via the Schwinger mechanism of MMs with Dirac charges 1$g_D$ $\leq$ $g$ $\leq$ 3$g_D$ and masses up to 75 GeV/c$^2$ at the 95\% confidence level. This is the first limit from a collider experiment to be based on a nonperturbative calculation of the MM production cross section. In addition, this is the first direct search that is sensitive to MMs that are not point-like. The current study constitutes therefore the first search in which finite-size MMs are potentially detectable, not subject to the exponential suppression of their production cross section expected in Drell-Yan or photon-fusion production. 

In the previous literature, exclusion mass limits spanning the 1725 -- 2370~GeV/c$^2$ range were obtained assuming Drell-Yan or photon-fusion production for various spin assumptions and magnetic charges up to 5$g_D$~\cite{atlas_2019,Dyons:2020}, which constrain light-monopole solutions of several theoretical models. The exclusion mass limits obtained here are below the TeV range because the production cross section is suppressed by the exponential of the MM mass. Nonetheless, the current search is pioneering due to its use of nonperturbatively calculated cross sections and is, to the best of our knowledge, the first to have sensitivity to composite monopole production at a collider. In order to ensure the soundness of our lower mass bounds, we have consistently made conservative assumptions. There is significant scope for future work to improve the accuracy of theoretical calculations, and for future experimental searches at higher energy and luminosity, so as to extend the mass reach.
The exclusion limits from the current search may guide both theorists and experimentalists in their quest towards understanding the nature of magnetic charge.

\begin{methods}
\label{sec:methods}

\subsection{Monte Carlo simulation of the MoEDAL experiment}

The MM simulation code is developed in \textsc{Gauss}~\cite{gauss}, which is the LHCb simulation framework that uses \textsc{Geant4} as the simulation engine. MoEDAL simulations use a dedicated \textsc{Geant4} class that describes production and propagation of MMs~\cite{MattKing_NPPP}. The MM ionization energy losses, geometry and material content of the MoEDAL detector and its vicinity are modeled in the simulation. The MMTs are described in \textsc{Geant4} as sensitive detectors and produce hits when MMs are trapped in them. These hits are recorded in simulation and analyzed for calculating efficiency and the expected rate of MMs detection. A custom-made momentum distribution of MMs derived from Schwinger kinematics is implemented and propagated through the MoEDAL geometry.

\subsection{Cross section calculation}
The electromagnetic fields are approximated as coherent, classical, event-averaged fields sourced by the heavy ions~\cite{Kharzeev:2007jp,Gursoy:2014aka}. In this approximation the fields are determined by integrating the Li\'{e}nard-Wiechert potentials over the charge distribution of the ions, which is modelled as a Woods-Saxon distribution with radius \(R_{\rm Pb}=6.62\)~fm and surface thickness \(a_{\rm Pb}=0.546\)~fm~\cite{Adam:2014qja}. We focus on ultraperipheral collisions, in which the fraction of colliding nucleons is small~\cite{Adam:2014qja}, and so do not attempt to model the effect of nucleon collisions on the field. Note that the inclusion of event-by-event fluctuations~\cite{Deng:2012pc} can be expected to significantly enhance the overall monopole yield, due to the exponential form of the cross sections, shown below. We leave this effect as a promising avenue for future work.

In the near vicinity of the collision center, the magnetic field is the largest component of the electromagnetic field, and it points perpendicularly to both the beam axis and the impact parameter. The electric field cancels at the collision centre but in the near vicinity it has a nonzero component parallel to the impact parameter: see Ref.~\cite{Huang_2016} for a review.

Following Ref.~\cite{arttu_prd2019}, we calculate the electromagnetic fields on a lattice of points in the vicinity of the collision centre, and perform a fit to the results. For lead-ion collisions at $5.02~\mathrm{TeV}$/nucleon, we find the peak magnetic field strength is \(B=7.6\pm 0.3~{\rm GeV}^2\), from which the field decays with inverse decay time \(\omega=73\pm 3\)~GeV, where we have used natural units, \(c=\hbar=\epsilon_0=1\). The uncertainty quoted on these values reflects the stability of the fit results with respect to the underlying assumptions of the fit model. In particular, some model assumptions put more weight on the agreement between the data and the theoretical computation in the region around the collision origin, while others fit the asymptotic tails better. Since we do not know which aspects of the data are most important for MM pair production, we take the differences between the fits as a measure of the uncertainty and assume a uniform distribution within this range. 
The magnitude of the peak magnetic field is in good agreement with Ref.~\cite{Deng:2012pc}, in which a Glauber-Model Monte-Carlo was used to model event-by-event fluctuations and nucleon collisions.


In Ref.~\cite{arttu_prd2019}, the monopole production cross section for fixed impact parameter $b$ was calculated in the LCFA and FPA approximations,
\begin{align*}
\frac{1}{2\pi b}\frac{d\sigma_{\rm LCFA}}{db} &\approx \frac{[g B(b)]^{4}}{18 \pi^3 M^4 \omega^2(b)\Omega^2(b)}\exp\left(-\frac{\pi M^2}{g B(b)}+\frac{g^2}{4}\right), \\
\frac{1}{2\pi b}\frac{d\sigma_{\rm FPA}}{db} &\approx \frac{[g B(b)]^{4}}{18 \pi^3 M^4 \omega^2(b)\Omega^2(b)} \exp\left(-\frac{4 M}{\omega(b)}\right),
\end{align*}
where $\Omega^2(b)\approx 2/(b R_{\rm Pb})$ is the product of the slow inverse decay lengths of the field perpendicular to the beam axis. Within these approximations, these equations are accurate up to a multiplicative \(O(1)\) factor. This uncertainty should have only a small effect on our final MM mass bounds due to the strong (exponential) dependence of the cross sections on the MM mass.

The total cross section in the LCFA is strongly dominated by impact parameters in the vicinity of $b=b_{\rm max}\approx 2R_{\rm Pb}$, at which the magnetic field is maximal.
As a consequence, integration over the impact parameter can be carried out using Laplace's method, yielding
\begin{equation}
\label{equ:LCFA}
\sigma_{\rm LCFA} \approx \frac{2(g B)^{9/2}R_{\rm Pb}^4}{9 \pi^2 M^5 \omega^2}\exp\left(-\frac{\pi M^2}{g B}+\frac{g^2}{4}\right),
\end{equation}
where all parameters are evaluated at $b=b_{\rm max}$. For the monopole masses we have studied, the range of impact parameters that contribute significantly to the total cross section deviates from $b_{\rm max}$ by less than about 1\%.

The total cross section in the FPA is also dominated by impact parameters in the vicinity of $b_{\rm max}$, though less strongly so. The integration over impact parameters yields
\begin{equation}
\label{equ:FPA}
\sigma_{\rm FPA} \approx \left[\frac{\omega}{M} + c_{\rm central}\right]\frac{2(gB)^4 R_{\rm Pb}^4}{9 \pi ^2 M^4 \omega ^2} \exp\left(-\frac{4 M}{\omega}\right),
\end{equation}
where all parameters are evaluated at $b=b_{\rm max}$. In this expression, the first term inside the square brackets arises from the contribution due to ultraperipheral collisions, with $b\geq b_{\rm max}$, and the second term $c_{\rm central}>0$ arises from integration over the contribution due to more central collisions with  $b< b_{\rm max}$.
Ignoring the effect of nucleon collisions on the production process results in $c_{\rm central}\approx 2/7$.
However, in our analysis we make the conservative choice $c_{\rm central}=0$, thereby avoiding altogether the complications of the conducting quark-gluon plasma and its effect on the electromagnetic fields, which remain an important source of uncertainty in predictions of the chiral magnetic effect~\cite{Tuchin:2013apa,Inghirami:2019mkc}.


Both the FPA and the LCFA are semiclassical approximations, yielding exponential forms for the cross section. For sufficiently light monopoles, e.g., \(M\lesssim 20\)~GeV in the FPA, the magnitude of the exponent is less than one, indicating the breakdown of the semi-classical approximation and consequently unsuppressed MM production.

For the distribution of momentum $\mathbf{p}$ we use the following result, based on the FPA, for the relative probability~\cite{Gould:2021bre}
\begin{equation}
\label{equ:relFPA}
\frac{d\sigma_{\rm FPA}(\mathbf{p})}{d\mathbf{p}}\Bigg/\frac{d\sigma_{\rm FPA}(\mathbf{0})}{d\mathbf{p}}= \exp\left[-\frac{4}{\omega} \left(\sqrt{M^2 + |\mathbf{p}|^2}-M\right) \right].
\end{equation}
Within the FPA, there is a residual \(O(1)\) uncertainty regarding the $p_z$-dependence which is not fully accounted for by this expression. This will result in an \(O(1)\) uncertainty in the final efficiency, which is subdominant compared to the uncertainty in the total cross section.

\subsection{Efficiency calculation} 
The trapping efficiency depends on the kinematics predicted by the MM model, the MM mass and magnetic charge, the material traversed by the MMs, and the corresponding energy losses. The efficiency is negligible for MMs with very low momentum, as slow-moving MMs would lose their energy through ionization before reaching the detectors. The efficiency increases almost linearly with increase in momentum until the MMs become energetic enough to pass through the MMTs without getting trapped. The efficiency then decreases with momentum above a threshold value that depends on the MM mass and magnetic charge. The Bethe-Bloch equation modified for MMs with magnetic charge $g$ = $n\cdot g_D$ ($n$ = 1, 2, 3...) that describes the ionization energy losses is given by:
 \begin{eqnarray*}
			\frac{dE}{dx} = \frac{4\pi e^{2} g^{2}}{m_{e}c^{2}} n_{e} \left[ \frac{1}{2} \ln{\frac{2 m_{e} c^{2} \beta^{2} \gamma^{2} T_{max}}{2I_{m}^{2}}} - \frac {K(|g|)}{2} - B(|g|) - \frac{\delta}{2}\right] \, ,
\end{eqnarray*}
where $n_{e}$ is the number of electrons per unit volume in the medium, $I_{m}$ is the mean ionization potential for MMs, and $K(|g|)$ = (0.406; 0.346 if $g > 1g_D$), $B(|g|)$ = (0.248, 0.672, 1.022, 1.243, 1.464) are correction terms for $g$ = $1g_D, ..., 5g_D$, respectively~\cite{Ahlen:1978}. Therefore, MMs with higher magnetic charges ($|g|$ $>$ 1$g_{D}$) are expected to have greater ionization losses and, as a consequence, tend to range out before reaching the MMTs, predominantly stopping in the upstream material. Also, keeping other parameters unchanged, MMs with lower masses have a higher velocity, leading to more energy losses, and hence failure to reach the MMTs. 

\subsection{Uncertainties and statistical methods}
The mean Poisson rate, $R_{exp}$, gives the expected number of MMs trapped in the MMTs during the MoEDAL Pb-Pb  data-taking run. It is calculated as the product of the MM cross section, the luminosity (exposure), and the trapping efficiency. We exclude MM masses with non-zero expected events, based on the nonobservation of MM after scanning the MMTs. The confidence level of exclusion (CL$_{exc}$) is determined from the Poisson statistics on $R_{exp}$, MC statistical errors, and systematic uncertainties on the detector geometry, cross section (due to $B$ and $\omega$ uncertainties), energy losses, and magnetometer response (false negatives, discussed in the following section). The MC statistical uncertainties are confined to less than 2 (7) percent for 1 (2) $g_D$ due to the large MC samples generated ($\sim$5$\cdot$10$^7$--1$\cdot$10$^{10}$, depending on the MM mass and magnetic charge) but increase to $\sim$20\% for 3$g_D$ due to the low trapping efficiency ($\approx$2.3$\cdot$10$^{-9}$). The false negative probability of the magnetometer response is 0.2\%~\cite{moedal_prl2019}. The $dE/dx$ calculation results in a relative uncertainty in the range 1-7\%~\cite{moedal_first}. The uncertainty of $B$ and $\omega$ is specified in the earlier Section. The remaining, dominant source of uncertainty is due to the material budget. The error in the measurements of component materials present in the VELO cavern is estimated by minimal and maximal geometry models, which encompass the uncertainties. Generally, the VELO vacuum vessel is modelled with high precision, while the cables and pipes present downstream of the VELO, as well as the inner regions of the vacuum pump and vacuum manifold are not as well known. We model the cables and small pipes as a set of vertical stainless-steel rods, which represent 2.3\% of the radiation length on average. We vary the rods’ radius in the simulation from 0.01 cm (minimal geometry) to 0.5 cm (maximum geometry). We also add a layer of stainless steel to the inner region of the vacuum tank. The default geometry contains 2.64 mm of stainless steel, representing 15\% of added radiation length. The minimal geometry contains no extra stainless steel. The maximal geometry, included as a conservative case, consists of 5.28 mm of extra stainless steel, i.e., 30\% of a radiation length.

Figure~\ref{fig:meanexpectedMMsFPA} shows an example of the dependence of the mean rate on the MM mass. The rate is calculated for the FPA cross section, which gives the conservative limits used in the Results section. Tables~\ref{tab:nexpFPA1} \textendash~\ref{tab:nexpFPA5} show the efficiency and expected rate of trapped MMs produced by the Schwinger mechanism for different values of the MM mass and magnetic charge. Cross section values are also shown, with a spread corresponding to $B$ and $\omega$ uncertainties. The spread in efficiency is between the maximal and minimal geometry models. We note that, while the resulting spread of $R_{exp}$ is large, its impact on the mass limits is strongly suppressed due to the exponential dependence of the cross section on the MM mass. 

The efficiency and cross section uncertainties lead to a systematic spread in $R_{exp}$. The probability to observe no MMs in the scanned MMTs given the range of $R_{exp}$, $P(0 | (R_{exp}^{min}, R_{exp}^{max}))$, is calculated using the toy MC approach. A uniform probability distribution for the range of possible efficiency values is assumed, reflecting no preference between the minimal and maximal geometry models. For $g = 3g_D$, the range's left boundary is the upper limit on the efficiency for the maximal geometry. Likewise, a uniform distribution is assumed for the $B$ and $\omega$ variables, which are expected to be be 100\% positively correlated. A toy value of the expected rate $R_{exp}^{i}$ is calculated using efficiency and cross section values based on random draws from their distributions. Then, a toy observed rate $R^{i}$ is randomly drawn from a Poisson distribution with $\mu = R_{exp}^{i}$. The process is repeated 10$^5$ times. The probability $P(0 | (R_{exp}^{min}, R_{exp}^{max}))$ is then defined as the fraction of $R^i=0$ entries in the toy $R^i$ distribution. Finally, the CL$_{exc}$ is calculated as $(1-P(0 | (R_{exp}^{min}, R_{exp}^{max})))\cdot100$\%. 

In order to test the robustness of this approach against the choice of the distribution of the efficiency, $\epsilon$, additional choices are considered. In the first test, a Gaussian distribution with $\mu=(\epsilon_{max} + \epsilon_{min})/2$ and $\sigma = (\mu-\epsilon_{min})/2$, truncated at $\epsilon_{min}$ and $\epsilon_{max}$, is assumed. In the second test, a skewed Gaussian distribution centered at $\epsilon$ corresponding to the default geometry, with $\sigma = (\mu-\epsilon_{min)}/2$ and skewness parameter set to 1 or 2, is assumed. The choice of a Gaussian distribution reflects our expectation that the default geometry is the most likely one. The direction of the skewness is dictated by the fact that the efficiency changes more rapidly when the material is removed from the geometry, compared to when it is added. In all considered cases the resulting CL$_{exc}$ agrees with or is slightly stronger than that for the uniform distribution of $\epsilon$. Additionally, for $g = 3g_D$, the calculation was repeated using uniform distribution with the maximal geometry's efficiency assumed to be two orders of magnitude lower than the limit obtained with the finite MC statistics (Tables~\ref{tab:nexpFPA3} \textendash~\ref{tab:nexpFPA5}). The mass limit decreases by $\sim$5 GeV/c$^2$.

\subsection{Magnetometer measurements} 
The MoEDAL trapping detector array comprising 2400 samples was exposed to 5.02~{\rm TeV}/nucleon LHC Pb-Pb collisions during the 2018 run. These samples were scanned for the presence of magnetic charges with a DC SQUID long-core magnetometer. Each sample was scanned through the superconducting coil at least twice, and the magnetometer's response was recorded before, during, and after each passage. The current induced in the superconducting coil is directly proportional to the difference in magnetic flux in the direction of transport. The magnetometer feedback, after multiplication by a calibration constant $C$ is converted into a magnetic pole strength $S$ (in units of $g_{D}$) . Two independent methods, the solenoid method and the convolution method were used to perform calibration of the instrument~\cite{DeRoeck2012}. A magnetic charge present in the sample would induce a persistent current in the superconducting coil, proportional to its pole strength. The difference between the measured induced currents before ($I_1$) and after ($I_2$) passage of the sample, while simultaneously adjusting for corresponding contributions ($I_1^{tray}$ and $I_2^{tray}$) of the empty conveyer tray, is defined as the persistent current. We calculate the magnetic pole strength as, 
\[S = C[(I_{2}-I_{1})-(I_{2}^{tray}-I_{1}^{tray})]\]
For a dipole, the currents induced by the north and south pole in a sample cancel out. The current expected from a Dirac MM was emulated using a long solenoid. A MM present in the sample would record a persistent current significantly different from zero. After passing through the superconducting coil, a sample was considered a MM candidate whenever the measured pole strength diverged from zero by more than 0.4 $g_{D}$. A number of 87 out of 2400 samples were identified as candidates. These candidates were scanned four or five times each. A sample with an actual magnetic charge would yield the same persistent current in repeated measurements. However, after multiple measurements on samples, whenever the first one diverged from zero, subsequent measurements were consistent with zero. The measured outliers could be attributed to spurious flux jumps occurring as a result of ferromagnetic impurities in the sample, noise currents in the superconducting loop, variations in external magnetic fields, and other known instrumental and environmental factors ~\cite{moedal_first}. An estimated false negative probability of less than 0.2\% is obtained for magnetic charges of 1 $g_{D}$. Thus, the presence of a MM with $\mid g \mid$ $\geq$ 0.5 $g_{D}$ was excluded in all samples at more than 99.75\% C.L.

\end{methods}

\newpage
\section*{Extended Data}

\begin{figure}[htpb]
\centering
\includegraphics[width=0.475\textwidth]{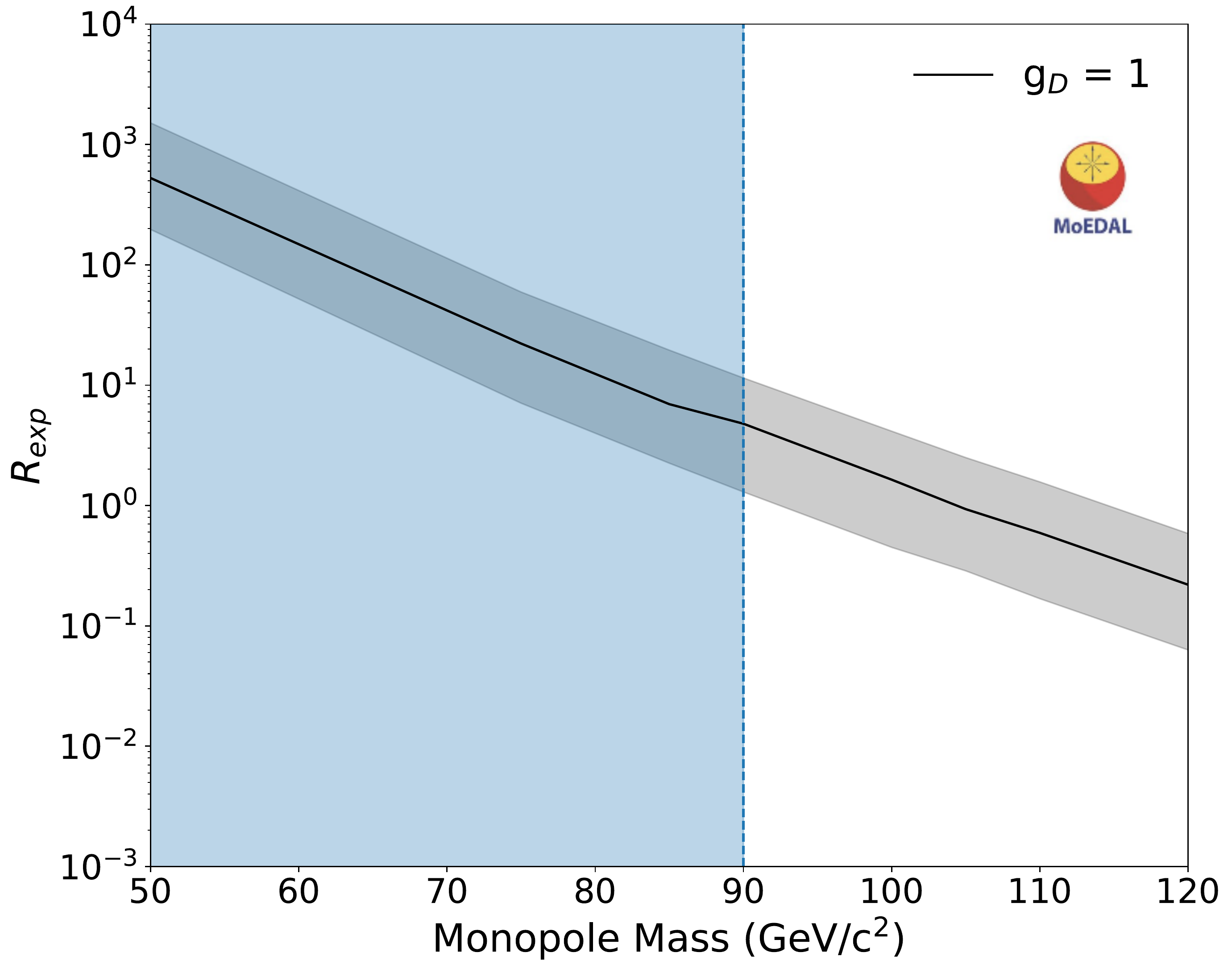}
\includegraphics[width=0.475\textwidth]{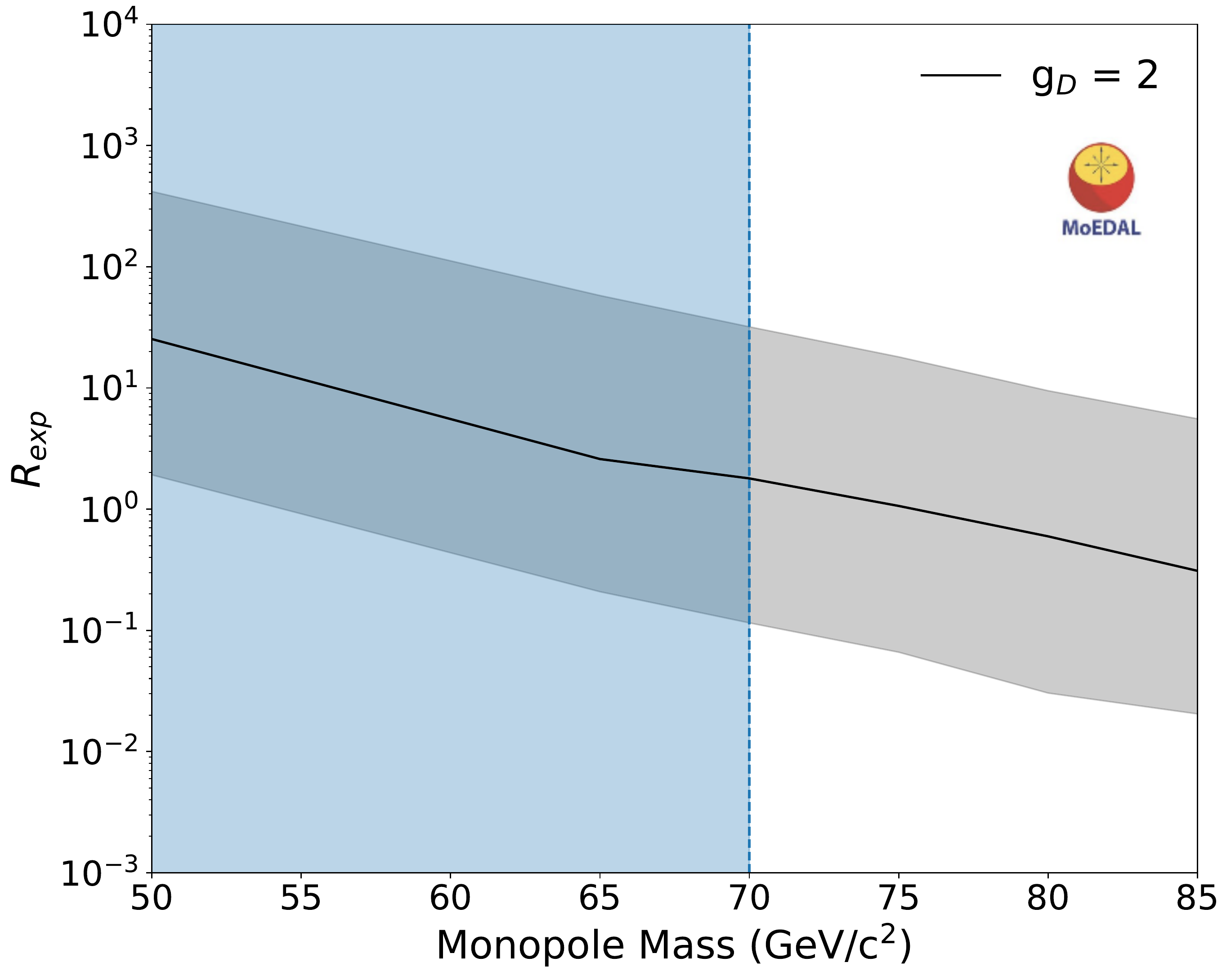}
\caption{The mean expected rate of MMs with 1 $g_D$ and 2 $g_D$ magnetic charge in the MMT as a function of the MM mass in the FPA model. The black line corresponds to the default geometry. The grey region corresponds to the systematic error, which is dominated by the material budget. The 95\% C.L. mass exclusion region is shown in blue.}
\label{fig:meanexpectedMMsFPA}
\end{figure}

\begin{table}[htpb]
\centering
\caption{Expected rate of MM trapping in the MoEDAL MMTs - $1g_D$ FPA model}
\label{tab:nexpFPA1}
\begin{tabular}{c|c|c|c|c}
 \hline
 
 $M$ (GeV/c$^2$) & $\mid g\mid$ ($g_D$) & $\sigma$ ($\mu$b) & $\epsilon$ $\cdot$10$^{-4}$ & $R_{exp}$ (min-max) \\
 \hline \hline
 50 & \multirow{ 6}{*}{1} & $(2.0^{+0.5}_{-0.4})\cdot$10$^4$  & 1.19$^{+1.46}_{-0.64}$ & $1.97\cdot$10$^{2}$--$1.50\cdot$10$^{3}$ \\
 \cline{1-1}\cline{3-5}
 75 & & $(6.6^{+2.0}_{-1.7})\cdot$10$^2$ & 1.34 $^{+1.61}_{-0.72}$ & $7.13$--$59.5$ \\
 \cline{1-1}\cline{3-5}
 80 & & $(2.0^{+0.7}_{-0.6})\cdot$10$^2$ & 1.41 $^{+1.66}_{-0.76}$ & $2.23$--$19.5$ \\
 \cline{1-1}\cline{3-5}
 85 & & $(1.2^{+0.3}_{-0.3})\cdot$10$^2$ & 1.45 $^{+1.68}_{-0.78}$ & $1.30$--$11.5$ \\
 \cline{1-1}\cline{3-5}
 100 & & $(4.0^{+1.5}_{-1.2})\cdot$10$^1$ & 1.52 $^{+1.57}_{-0.83}$ & $0.45$--$4.16$ \\
 \cline{1-1}\cline{3-5}
 105 & & $(2.4^{+0.9}_{-0.8})\cdot$10$^1$ & 1.57 $^{+1.68}_{-0.82}$ & $0.29$--$2.51$ \\
 \hline 
 \end{tabular}
\end{table}

\begin{table}[htpb]
\centering
\caption{Expected rate of MM trapping in the MoEDAL MMTs - $2g_D$ FPA model}
\label{tab:nexpFPA2}
\begin{tabular}{c|c|c|c|c}
 \hline
 $M$ (GeV/c$^2$) & $\mid g\mid$ ($g_D$) & $\sigma$ ($\mu$b) & $\epsilon$ $\cdot$10$^{-7}$ & $R_{exp}$ (min-max) \\
 \hline \hline
 50 & \multirow{6}{*}{2} & $(3.1^{+0.8}_{-0.7})\cdot$10$^5$ & 3.3 $^{+43}_{-3.0}$ & $1.9$--$4.1\cdot$10$^{2}$ \\
 \cline{1-1}\cline{3-5}
 65 & & $(3.7^{+1.0}_{-0.9})\cdot$10$^4$ & 3.8 $^{+48}_{-3.5}$ & $0.2$--$57$ \\
 \cline{1-1}\cline{3-5}
 70 & & $(2.0^{+0.6}_{-0.5})\cdot$10$^4$ & 3.8 $^{+50}_{-3.5}$ & $0.1$--$32$ \\
 \cline{1-1}\cline{3-5}
 75 & & $(1.1^{+0.3}_{-0.3})\cdot$10$^4$ & 4.5 $^{+51}_{-4.2}$ & $0.1$--$18$ \\
 \cline{1-1}\cline{3-5}
 80 & & $(5.8^{+1.8}_{-1.6})\cdot$10$^3$ & 4.3 $^{+49}_{-4.0}$ & $0.0$--$9.5$ \\
 \cline{1-1}\cline{3-5}
 85 & & $(3.3^{+1.1}_{-0.9})\cdot$10$^3$ & 4.4 $^{+50}_{-4.1}$ & $0.0$--$5.6$ \\
 \hline 
\end{tabular}
\end{table}


\begin{table}[htpb]
\centering
\caption{Expected rate of MMs trapping in the MoEDAL MMTs - $3g_D$ FPA model}
\label{tab:nexpFPA3}
\begin{tabular}{c|c|c|c|c}
 \hline
 $M$ (GeV/c$^2$) & $\mid g\mid$ ($g_D$) & $\sigma$ ($\mu$b) & $\epsilon_{min-max}$ & $R_{exp}$ (min-max) \\
 \hline \hline
 50 & \multirow{ 6}{*}{3} & $(1.6^{+0.4}_{-0.4})\cdot$10$^6$ & $<1\cdot$10$^{-10}$--$1.2\cdot$10$^{-6}$ & $0.0$--$5.3\cdot$10$^{2}$\\
 \cline{1-1}\cline{3-5}
 65 & & $(1.9^{+0.6}_{-0.5})\cdot$10$^5$ & $<1\cdot$10$^{-10}$--$1.3\cdot$10$^{-6}$ & $0.0$--$75$\\
 \cline{1-1}\cline{3-5}
 70 & & $(9.9^{+2.9}_{-2.5})\cdot$10$^4$ & $<1\cdot$10$^{-10}$--$1.2\cdot$10$^{-6}$  & $0.0$--$35$ \\
 \cline{1-1}\cline{3-5}
 75 & & $(5.3^{+1.6}_{-1.4})\cdot$10$^4$ & $<1\cdot$10$^{-10}$--$1.2\cdot$10$^{-6}$ & $0.0$--$19$\\
 \cline{1-1}\cline{3-5}
 80 & & $(2.9^{+0.9}_{-0.8})\cdot$10$^4$ & $<1\cdot$10$^{-10}$--$1.3\cdot$10$^{-6}$ & $0.0$--$11$\\
 \cline{1-1}\cline{3-5}
 85 & & $(1.7^{+0.6}_{-0.5})\cdot$10$^4$ & $<1\cdot$10$^{-10}$--$1.3\cdot$10$^{-6}$ & $0.0$--$6.6$\\
 \hline
 \end{tabular}
\end{table}

\begin{table}[htpb]
\centering
\caption{Expected rate of MM trapping in the MoEDAL MMTs - $4g_D$ FPA model}
\label{tab:nexpFPA4}
\begin{tabular}{c|c|c|c|c}
 \hline
 $M$ (GeV/c$^2$) & $\mid g\mid$ ($g_D$) & $\sigma$ ($\mu$b) & $\epsilon_{min-max}$  & $R_{exp}$ (max-min) \\
 \hline \hline
 50 & \multirow{6}{*}{4} & $(5.0^{+1.2}_{-1.1})\cdot$10$^6$ & $<2\cdot$10$^{-9}$--$3.5\cdot$10$^{-7}$ & $0.0$--$5.1\cdot$10$^{2}$\\
 \cline{1-1}\cline{3-5}
 65 & & $(6.0^{+1.7}_{-1.5})\cdot$10$^5$ & $<2\cdot$10$^{-9}$--$3.9\cdot$10$^{-7}$ & $0.0$--$69$ \\
 \cline{1-1}\cline{3-5}
 70 & & $(3.1^{+0.9}_{-0.8})\cdot$10$^5$ & $<2\cdot$10$^{-9}$--$3.4\cdot$10$^{-7}$ & $0.0$--$32$ \\
 \cline{1-1}\cline{3-5}
 85 & & $(1.7^{+0.5}_{-0.4})\cdot$10$^5$ & $<2\cdot$10$^{-9}$--$2.8\cdot$10$^{-7}$ & $0.0$--$14$ \\
 \cline{1-1}\cline{3-5}
 80 & & $(9.3^{+2.9}_{-2.5})\cdot$10$^4$ & $<2\cdot$10$^{-9}$--$3.6\cdot$10$^{-7}$ & $0.0$--$10$\\
 \cline{1-1}\cline{3-5}
 85 & & $(5.2^{+1.7}_{-1.5})\cdot$10$^4$ & $<2\cdot$10$^{-9}$--$3.6\cdot$10$^{-7}$ & $0.0$--$5.8$ \\
 \hline
 \end{tabular}
\end{table}

\begin{table}[htpb]
\centering
\caption{Expected rate of MM trapping in the MoEDAL MMTs - $5g_D$ FPA model}
\label{tab:nexpFPA5}
\begin{tabular}{c|c|c|c|c}
 \hline
 $M$ (GeV/c$^2$) & $\mid g\mid$ ($g_D$) & $\sigma$ ($\mu$b) & $\epsilon_{min-max}$  & $R_{exp}$ (max-min) \\
 \hline \hline
 50 & \multirow{6}{*}{5} & $(1.2^{+0.3}_{-0.3})\cdot$10$^7$ & $<2\cdot$10$^{-9}$--$1.0\cdot$10$^{-7}$ & $0.0$--$3.7\cdot$10$^{2}$ \\
 \cline{1-1}\cline{3-5}
 60 & & $(2.9^{+0.8}_{-0.7})\cdot$10$^6$ & $<2\cdot$10$^{-9}$--$1.1\cdot$10$^{-7}$ & $0.0$--$93$ \\
 \cline{1-1}\cline{3-5}
 65 & & $(1.5^{+0.4}_{-0.4})\cdot$10$^6$ & $<2\cdot$10$^{-9}$--$1.4\cdot$10$^{-7}$ & $0.0$--$60$ \\
 \cline{1-1}\cline{3-5}
 70 & & $(7.6^{+2.2}_{-1.9})\cdot$10$^5$ & $<2\cdot$10$^{-9}$--$1.1\cdot$10$^{-7}$ & $0.0$--$25$ \\
 \cline{1-1}\cline{3-5}
 75 & & $(4.1^{+1.3}_{-1.1})\cdot$10$^5$ & $<2\cdot$10$^{-9}$--$1.0\cdot$10$^{-7}$ & $0.0$--$12$ \\
 \cline{1-1}\cline{3-5}
 80 & & $(2.3^{+0.7}_{-0.6})\cdot$10$^5$ & $<2\cdot$10$^{-9}$--$0.9\cdot$10$^{-7}$ & $0.0$--$6.4$ \\
 \hline
\end{tabular}
\end{table}

\newpage
\bibliography{ref.bib}

\begin{addendum}
 \item We thank CERN for the LHC's successful Run-2 operation, as well as the support staff from our institutions without whom MoEDAL could not be operated. We acknowledge the invaluable assistance of particular members of the LHCb Collaboration: G. Wilkinson, R. Lindner, E. Thomas and G. Corti. Computing support was provided by the GridPP Collaboration, in particular by the Queen Mary University of London and Liverpool grid sites. This work was supported by grant PP00P2 150583 of the Swiss NSF; by the UK Science and Technology Facilities Council, via the grants, ST/L000326/1, ST/L00044X/1, ST/N00101X/1, ST/P000258/1, ST/P000762/1,
ST/T000759/1 and ST/T000791/1; by the Generalitat Valenciana via a special grant for MoEDAL and via the projects PROMETEO-II/2017/033 and PROMETEO/2019/087; by MCIU / AEI / FEDER, UE via the grants FPA2016-77177-C2-1-P, FPA2017-85985-P, FPA2017-84543-P and PGC2018-094856-B-I00; by the Physics Department of King's College London; by NSERC via a project grant; by the V-P Research of the University of Alberta (UofA); by the Provost of the UofA); by UEFISCDI (Romania); by the INFN (Italy); by the Estonian Research Council
via a Mobilitas Pluss grant MOBTT5; and by the NSF grant 2011214 to the University of Alabama MoEDAL group. A.R.~was also supported by Institute for Particle Physics Phenomenology Associateship.
 
 \item[Author Contributions] The Monopole and Exotics Detector at the LHC was constructed and is maintained by the MoEDAL collaboration. A large number of authors contributed to the data processing,
detector calibration and MC simulations used in this work. The MoEDAL collaboration
acknowledges the substantial contributions to this manuscript from A.U., I.O., A.R., D.H., O.G., N.M., and J.P. The final manuscript was reviewed and approved by all authors.
 
 \item[Competing Interests] The authors declare that they have no
competing financial interests.
 \item[Correspondence] Correspondence and requests for materials
should be addressed to I. Ostrovskiy~(email: iostrovskiy@ua.edu).
\item[Data Availability] The SQUID magnetometer data are stored at the ETH Zurich computers, with copies on the cloud. The MC data are stored at the CERN lxplus server. Inquiries regarding access to the data may be directed to the corresponding author.
\item[Code Availability] The MoEDAL collaboration develops and maintains the code used for the simulation of the experimental apparatus and statistical analysis of the raw data used in this result. This code is shared among the collaboration, but not publicly distributed. Inquiries regarding the algorithms and methods used in this result may be directed to the corresponding author.
\end{addendum}

\end{document}